# Bulk Viscous Bianchi Type-$V$ Cosmological Models with Time Function $G$ and $\Lambda$

Nawsad Ali

*Department of Mathematics,*

*West Goalpara College, Balarbhita-783129, Goalpara, Assam, India*

Email:  alinawsad@rocketmail.com

**Abstract**: In this paper we are to study homogeneous and anisotropic Bianchi type-$V$ universe in presence of viscous fluid as source of matter with time function gravitational constant $G$ and cosmological term $\Lambda$. The viscosity coefficient is regarded as a power function of matter density in the first case whereas in other case it is considered as proportional to scale of expansion. Physically realistic solutions of the field equations are obtained by using a fractional form of Hubble's parameter, which leads an early deceleration and late time acceleration of the universe. Certain physical and geometrical behaviors of the models are also studied.

## 1. Introduction

In general relativity, the Newtonian gravitational constant $G$ and cosmological term $\Lambda$ are supposed to be fundamental constants. The gravitational constant $G$ displays the role of coupling constant between space geometry and matter in Einstein's field equations. But in an expanding universe, it is natural to consider as a function of cosmic time. At first Dirac [1-2] and Dicke [3] investigated the possibility of time variable $G$. Different types of hypothesis investigated by Dirac that leads to a cosmology with time variable $G$. In addition, Canuto et al. [4] have given numerous explanations on the basis of several arguments that $G$ is time dependent. Furthermore, Canuto and Narlikar [5] have suggested that time function $G$ cosmology consistent with the cosmological observations available at literature. At recent past time varying $G$ in general relativity have been studied by Levitt [6]; Beesham [7] and Abdel-Rahman [8].

Similarly, the cosmological term $\Lambda$ is taken to be one of the most essential unsolved phenomena in relativistic cosmology. Earlier this cosmological constant $\Lambda$ was considered to describe the static solution to the Einstein's field equation. But the recent work in ref. [9-12] described that a tiny positive value of $\Lambda$, as one of the theoretical candidate of dark energy which cause the accelerate expansion of the universe. In the study of relativistic quantum field theory the cosmological term $\Lambda$ interpreted as the energy density of the vacuum [13-15], whereas Linde [16] has studied that the cosmological term $\Lambda$ is a variable of temperature and is related with the process of broken symmetries. Hence, it could be time dependent in expanding universe [17]. Cosmological scenarios with time variable $\Lambda$ were investigated many authors [18-22].

The linking of the time variable $G$ with that of $\Lambda$ in the relativistic framework was proposed by Lau [23]. This new concept leaves field equations formally unchanged as time function of $\Lambda$ is together with the time variable of $G$. Beesham [24] and Kaligas et al. [25] have investigated linking the time variable of $G$ together with that of $\Lambda$ in general theory relativity. Recently, cosmological models with time dependent $G$ and $\Lambda$ term have been analyzed in a series of works (Vishwakarma [26], Abdussatter and Vishwakarma [27]; Beesham [28]; Saha [29]; Yadav et al. [30]; Singh et al. [31].

In general cosmological models usually have considered to the cosmic fluid as a perfect fluid. However, these models need not to incorporate the dissipation mechanism as responsible for initial anisotropies. The bulk viscosity appears any time a fluid expands and consequently it ceases to be in thermodynamic equilibrium. Thus, the bulk viscosity is appraise of the pressure needed to restore balance to a compressed form of expanding system. Additionaly, it is think that during neutrino decoupling, the matter behaved like as viscous fluid in the early era of evolution (Klimek [32]. Murphy [33] has explained the role of viscosity in avoiding the initial singularity whereas Padmanabhan and Chitre [34] have suggested that bulk viscosity leads to an inflationary like solution. Bulk viscous models with time varying $G$ and $\Lambda$ terms have been studied by Arbab [35]; Beesham et al. [36]; Bali and Tinker [37], where viscous coefficient chosen as a power function of matter density.' Recently, Baghel and Singh [38] have explained Bianchi type-$V$ bulk viscous universe with varying $G$ and $\Lambda$ terms, where they considered the viscous coefficient as a linear function of Hubble parameter. Addition viscous coefficient as a quadratic function of Hubble parameter for axially symmetric Bianchi type-$I$ universe with time dependent $G$ and $\Lambda$ terms have analyzed by Das and Ali [39].

The anisotropy of cosmic expansion, which, as it is generally assumed, diminishes with time, is a very important quantity. The current experimental data and theoretical ideas (Chimento [40], Misner [41]), confirmed the existence of anisotropy expansion turning into isotropy expansion (Land and Magueijo [42]). It is this hypothesis that dictates the consideration of the observed models with anisotropic background. The Bianchi type-$V$ homogeneous models are the simplest anisotropic models, whose spatial sections are flat but the rate of expansion or contraction are dependent of direction. Roy and Singh [43, 44] have discussed Bianchi type-$V$ models with electromagnetic field. Banerjee and Sanyal [45] have investigated Bianchi type-$V$ cosmologies with viscosity and heat flow. Several authors viz. Beesham [46], Collins [47], Maartens and Nel [48], Roy and Prasad [49], Bali and Jain [50], Coley [51], Pradhan and Yadav [52], Singh and Baghel [53] and Yadav [54] have analyzed Bianchi type-$V$ models in different physical contexts.

In this paper we present anisotropic Bianchi type-$V$ universe is considered with time variable $G$ and $\Lambda$ in the presence of viscous fluid. The viscous coefficient assumed to be $\zeta = \zeta_0 \rho^l$ and $\zeta = \zeta_1 \theta$, where $\zeta_0$, $\zeta_1$ and $l$ are constants. To get theoretical solutions of the present model of the universe we consider fractional form of average Hubble's parameter, which depicts earlier deceleration and current time acceleration of the universe is explained by Ellis and Madsen [55].

This paper is organized as follows: the metric and field equations are presented in Section 2. In Section 3, we deal with a particular form of generalized Hubble's parameter, which gives an early deceleration and late time acceleration and the coefficient of bulk viscosity as a function of matter density and expansion scalar. In Section 3.1, we consider the coefficient of bulk viscosity $(\zeta)$ as power function of matter density $(\rho)$ whereas in Section 3.2 $\zeta$ is taken as proportional to scale of expansion $(\theta)$ of the model. Section 4 includes the physical discussion. In Section 5, we have given the concluding remarks

## 2. The Metric and Field Equations

We consider Bianchi type-$V$ space time in an orthogonal form is given by the line element

$$ds^2 = -dt^2 + X^2(t)dx^2 + e^{2\alpha x}\left\{Y^2(t)dy^2 + Z^2(t)dz^2\right\}, \qquad (1)$$

where $\alpha$ is a constant. We consider the cosmic matter consisting of bulk viscous fluid represented by the energy momentum tensor

$$T_{ij} = (\rho + \bar{p})v_i v_j + \bar{p} g_{ij}, \qquad (2)$$

with

$$\bar{p} = p - \zeta v^i_{;i}, \qquad (3)$$

Obeying the equation of state

$$p = \omega \rho, \qquad (4)$$

where $p$, $\rho$, $\zeta$ are respectively the equilibrium pressure, the energy density of matter, the coefficient of bulk viscosity and $v^i$ is the four velocity vector of the fluid satisfying $v_i v^i = -1$. The semicolon in equation (3) stands for covariant differentiation. On the ground of thermodynamic, the viscous coefficient $\zeta$ is positive, assuring that the viscosity can pushes the dissipative pressure $\bar{p}$ towards negative values.

The Einstein's field equations for time variable $G(t)$ and $\Lambda(t)$ are given by

$$R_{\mu\nu} - \frac{1}{2} g_{\mu\nu} R = -8\pi G(t) T_{\mu\nu} + \Lambda(t) g_{\mu\nu}. \qquad (5)$$

The Einstein's field equations (5) for the metric (1) becomes to

$$8\pi G \bar{p} - \Lambda = \frac{\alpha^2}{X^2} - \frac{\ddot{Y}}{Y} - \frac{\ddot{Z}}{Z} - \frac{\dot{Y}\dot{Z}}{YZ}, \qquad (6)$$

$$8\pi G \bar{p} - \Lambda = \frac{\alpha^2}{X^2} - \frac{\ddot{Z}}{Z} - \frac{\ddot{X}}{X} - \frac{\dot{Z}\dot{X}}{ZX}, \tag{7}$$

$$8\pi G \bar{p} - \Lambda = \frac{\alpha^2}{X^2} - \frac{\ddot{X}}{X} - \frac{\ddot{Y}}{Y} - \frac{\dot{Y}\dot{Z}}{XY}, \tag{8}$$

$$8\pi G \rho + \Lambda = -\frac{3\alpha^2}{X^2} + \frac{\dot{X}\dot{Y}}{XY} + \frac{\dot{Y}\dot{Z}}{YZ} + \frac{\dot{Z}\dot{X}}{ZX}, \tag{9}$$

$$0 = \frac{2\ddot{X}}{X} - \frac{\ddot{Y}}{Y} - \frac{\ddot{Z}}{Z}, \tag{10}$$

where an overhead dot represents the derivatives with respect to the field variable cosmic time $t$. The vanishing of divergence of Einstein tensor becomes to

$$8\pi G \left\{ \dot{\rho} + (\rho + \bar{p}) \left( \frac{\dot{X}}{X} + \frac{\dot{Y}}{Y} + \frac{\dot{Z}}{Z} \right) + \frac{\dot{G}}{G} \rho \right\} + \dot{\Lambda} = 0. \tag{11}$$

The energy conservation equation for bulk viscous fluid $T^{\nu}_{\mu;\nu} = 0$ may be obtain as

$$\dot{\rho} + (\rho + \bar{p}) \left( \frac{\dot{X}}{X} + \frac{\dot{Y}}{Y} + \frac{\dot{Z}}{Z} \right) = 0. \tag{12}$$

Equations (11) and (12) put $G$ and $\Lambda$ in some sort of couple field as

$$8\pi \dot{G} \rho + \dot{\Lambda} = 0. \tag{13}$$

From equation (13) we can say that, when $\Lambda$ is constant or $\Lambda = 0$, $G$ turns out to be constant for non zero value of energy density. The average scale factor $a$ for the metric (1) is defined as

$$a^3 = \sqrt{-g} = XYZ. \tag{14}$$

From equations (6)-(10) and (14), we obtain

$$\frac{\dot{X}}{X} = \frac{\dot{a}}{a}, \tag{15}$$

$$\frac{\dot{Y}}{Y} = \frac{\dot{a}}{a} - \frac{k_1}{a^3}, \tag{16}$$

and

$$\frac{\dot{Z}}{Z} = \frac{\dot{a}}{a} + \frac{k_1}{a^3}, \tag{17}$$

with $k_1$ is a constant of integration. Now integrating equations (15)-(17), we get

$$X = m_1 a, \tag{18}$$

$$Y = m_2 a \exp\left\{-k_1 \int \frac{dt}{a^3}\right\} \tag{19}$$

and

$$Z = m_3 a \exp\left\{k_1 \int \frac{dt}{a^3}\right\}, \tag{20}$$

where $m_1$, $m_2$ and $m_3$ are integrating constants with $m_1 m_2 m_3 = 1$.

We define the generalized mean Hubble parameter $H$ and the deceleration parameter $q$ as

$$H = \frac{\dot{a}}{a} = \frac{1}{3}\left(H_x + H_y + H_z\right), \tag{21}$$

$$q = -1 - \frac{\dot{H}}{H^2}, \tag{22}$$

where $H_x = \frac{\dot{X}}{X}$, $H_y = \frac{\dot{Y}}{Y}$ and $H_z = \frac{\dot{Z}}{Z}$ are directional Hubble's factors along $x$, $y$ and $z$ directions respectively.

We introduced the expansion scalar $\theta$ and shear scalar $\sigma$ as usual

$$\theta = v^{\mu}_{;\nu} \; ; \quad \sigma^2 = \frac{1}{2}\sigma_{\mu\nu}\sigma^{\mu\nu}, \tag{23}$$

$\sigma^{ij}$ being shear tensor and semicolon stand for covariant derivative. For Bianchi type-$V$ metric, expansion scalar $\theta$ and shear scalar $\sigma$ are as follows

$$\theta = 3\frac{\dot{a}}{a}, \tag{24}$$

$$\sigma = \frac{k_1}{a^3}. \tag{25}$$

Equations (6)-(10) and (12) can be explicitly written in terms of $H$, $\sigma$ and $q$ as

$$8\pi G \bar{p} - \Lambda = (2q-1)H^2 - \sigma^2 + \frac{\alpha^2}{a^2}, \tag{26}$$

$$8\pi G \rho + \Lambda = 3H^2 - \sigma^2 - \frac{3\alpha^2}{a^2}, \tag{27}$$

and

$$\dot{\rho} + 3(\rho + \bar{p})H = 0. \tag{28}$$

Note that energy density of the universe is a positive quantity. At the early times of the evolution when the average scale factor $a$ was closer to zero, the energy density of the universe was very large. Again, with the expansion of the universe i.e. with increase value of $a$, the energy density decreases and for very big value of $a$, $\rho \to 0$. In that case from (27), we obtain $\frac{\rho_v}{\rho_c} \to 1$, where $\rho_v = \frac{\Lambda}{8\pi G}$ and $\rho_c = \frac{3H^2}{8\pi G}$. For $\Lambda \geq 0$, $\rho \leq \rho_c$. Also, from (27) we obtain

$$\frac{\sigma^2}{\theta^2} = \frac{1}{3} - \frac{8\pi G \rho}{\theta^2} - \frac{3\alpha^2}{a^2 \theta^2} - \frac{\Lambda}{\theta^2}. \tag{29}$$

Therefore, $0 \leq \frac{\sigma^2}{\theta^2} \leq \frac{1}{3}$ and $0 \leq \frac{8\pi G \rho}{\theta^2} \leq \frac{1}{3}$ for $\Lambda \geq 0$. Thus the presence of a positive $\Lambda$ puts restriction on the higher limit of anisotropy wherein a negative $\Lambda$ contributes to the anisotropy. From equation (26) and (27), we obtain

$$\frac{d\theta}{dt} = \Lambda + 12\pi \theta \zeta G - 4\pi G(\rho + 3p) - 2\sigma^2 - \frac{1}{3}\theta^2, \tag{30}$$

which is the Raychaudhuri equation for viscous fluid distribution. We note that for $\Lambda \leq 0$ and $\zeta = 0$, the model will always be in decelerating stage provided the strong energy condition [56] satisfies. However, in presence of viscosity, the positive $\Lambda$ will slow down the rate of decrease of volume expansion. Also, $\dot{\sigma}=-\sigma\theta$ reduces that $\sigma$ decrease in an evolving universe and for an infinitely big value of $a$, $\sigma$ tends to zero.

## 3. Solution of the Field Equation

Since there are six linearly independent equations (4) and (6)-(10) with eight unknown field variables $X, Y, Z, p, \rho, G, \Lambda$ and $\zeta$. Hence the system is initially undetermined accompanied with the energy conservation equation (12), and we need two extra additional constraints to solve the system completely. Since the metric (1) is completely characterized by average scale factor $a$ and usually it is connected with the average Hubble parameter $H$. Following, Ellis and Madsen [55], firstly, we assumed a fractional form of average Hubble parameter $H$, which yields both decelerating and accelerating phase of the universe. The average Hubble parameter $H$ is simply related to the average scale factor $a$ of the anisotropic Bianchi type-$V$ universe as

$$H(a) = \beta(a^{-\gamma} + 1), \tag{31}$$

where $\beta(>0)$ and $\gamma(>1)$ are constants. For this consideration, the deceleration parameter $q$ turns out to be

$$q = \frac{\gamma}{a^\gamma + 1} - 1. \tag{32}$$

This types of mathematical expression $q$ has also been explained by many authors in refs. [57-61]. From equation (32), we observe that when $a=0$, $q=\gamma-1>0$; $q=0$ for $a^\gamma = \gamma-1$ and when $a^\gamma > \gamma-1$ then $q<0$. We choose the value $a=0$ for $t=0$. Thus, the universe begins with an initial expansion and it changes from decelerating stage to an accelerating one. This cosmological scenario is agreement with the latest supernova observation (SN 1a).

Secondly, to determined the coefficient bulk viscosity we assume $\zeta = \zeta_0 \rho^l$ and $\zeta = \zeta_1 \theta$, where $\zeta_0 (\geq 0)$, $\zeta_1$ and $l(\geq 0)$ are constants.

For our model average scale factor $a$ is given by

$$a^\gamma = e^{\beta\gamma(t+t_0)} - 1. \tag{33}$$

where $t_0$ is a constant of integration. Setting $a=0$ for $t=0$, we get $t_0 = 0$. Therefore,

$$a^\gamma = e^{\beta\gamma t} - 1. \tag{34}$$

### 3.1 Cosmology for $\zeta = \zeta_0 \rho^l$

To determine the coefficient of bulk viscosity $\zeta$ is assumed to be a power function of matter density $\rho$ [62-64], i.e.

$$\zeta = \zeta_0 \rho^l, \qquad (35)$$

For this choice, equation (28) becomes to

$$\dot{\rho} + 3(1+\omega)\rho H = 9\zeta_0 \rho^l H^2. \qquad (36)$$

Solving (31), (34) and (36) we get the energy density of matter field $\rho$ and the coefficient of bulk viscosity $\zeta$ as

$$\rho = \frac{3\beta\zeta_0}{(1+\omega)} + \frac{9\beta\zeta_0(1-l)}{\{3(1-l)(1+\omega)-\gamma\}(e^{\beta\eta}-1)} + \frac{k_2}{(e^{\beta\eta}-1)^{\frac{3(1-l)(1+\omega)}{\gamma}}}, \qquad (37)$$

$$\zeta = \zeta_0 \left[ \frac{3\beta\zeta_0}{(1+\omega)} + \frac{9\beta\zeta_0(1-l)}{\{3(1-l)(1+\omega)-\gamma\}(e^{\beta\eta}-1)} + \frac{k_2}{(e^{\beta\eta}-1)^{\frac{3(1-l)(1+\omega)}{\gamma}}} \right]^l, \qquad (38)$$

where $k_2$ is a constant of integration. The gravitational constant $G$ and cosmological term-$\Lambda$ are leads as

$$G = \frac{1}{4\pi} \left[ \frac{\beta^2 \gamma e^{\beta\eta}}{(e^{\beta\eta}-1)^2} - \frac{k_1^2}{(e^{\beta\eta}-1)^{\frac{6}{\gamma}}} - \frac{\alpha^2}{(e^{\beta\eta}-1)^{\frac{2}{\gamma}}} \right] \times \left[ 3\beta\zeta_0 + \frac{3\beta\zeta_0(1-l)(1+\omega)}{\{3(1-l)(1+\omega)-\gamma\}(e^{\beta\eta}-1)} \right.$$

$$\left. + \frac{k_2(1+\omega)}{(e^{\beta\eta}-1)^{\frac{3(1-l)(1+\omega)}{\gamma}}} - \frac{3\beta\zeta_0 e^{\beta\eta}}{(e^{\beta\eta}-1)} \left\{ \frac{3\beta\zeta_0}{(1+\omega)} + \frac{9\beta\zeta_0(1-l)}{\{3(1-l)(1+\omega)-\gamma\}(e^{\beta\eta}-1)} + \frac{k_2}{(e^{\beta\eta}-1)^{\frac{3(1-l)(1+\omega)}{\gamma}}} \right\}^{l} \right]^{-1},$$

$$(39)$$

$$\Lambda = \left[ \frac{2k_1^2}{(e^{\beta\eta}-1)^{\frac{6}{\gamma}}} + \frac{2\alpha^2}{(e^{\beta\eta}-1)^{\frac{2}{\gamma}}} - \frac{2\beta^2 \gamma e^{\beta\eta}}{(e^{\beta\eta}-1)^2} \right] \times \left[ \frac{3\beta\zeta_0}{(1+\omega)} + \frac{9\beta\zeta_0(1-l)}{\{3(1-l)(1+\omega)-\gamma\}(e^{\beta\eta}-1)} \right.$$

$$+\frac{k_2}{(e^{\beta\eta}-1)^{\frac{3(1-l)(1+\omega)}{\gamma}}}\Bigg]\times\Bigg[3\beta\zeta_0+\frac{3\beta\zeta_0(1-l)(1+\omega)}{\{3(1-l)(1+\omega)-\gamma\}(e^{\beta\eta}-1)}+\frac{k_2(1+\omega)}{(e^{\beta\eta}-1)^{\frac{3(1-l)(1+\omega)}{\gamma}}}$$

$$-\frac{3\beta\zeta_0 e^{\beta\eta}}{(e^{\beta\eta}-1)}\Bigg\{\frac{3\beta\gamma\zeta_0}{(1+\omega)}+\frac{9\beta\zeta_0(1-l)}{\{3(1-l)(1+\omega)-\gamma\}(e^{\beta\eta}-1)}+\frac{k_2}{(e^{\beta\eta}-1)^{\frac{3(1-l)(1+\omega)}{\gamma}}}\Bigg\}^{l}\Bigg]^{-1}$$

$$+\frac{3\beta^2 e^{2\beta\eta}}{\left(e^{\beta\eta}-1\right)^2}-\frac{k_1^{\;2}}{\left(e^{\beta\eta}-1\right)^{\frac{6}{\gamma}}}-\frac{3\alpha^2}{\left(e^{\beta\eta}-1\right)^{\frac{2}{\gamma}}}. \tag{40}$$

When $t=0$, the values of all expressions $\zeta$, $\rho$, $\Lambda$ and $G$ are infinite. For $t\to\infty$, $\zeta=\dfrac{3m\zeta_0^{\;2}}{(1+\omega)}$ and $\rho=\dfrac{3\beta\zeta_0}{(1+\omega)}$.

From equation (37) the matter density is a decreasing function of cosmic time (see Fig. 1). From equation (39) we have drawn $G$ against cosmic time $t$, which is negative. (see Fig. 2). The negative behavior of $G$ has studied by Starobinskii [65]. Also, from equation (40), it is clear that the cosmological term $\Lambda$ being big value at early times and relaxes to a small value at the current time (see Fig. 3).

### 3.2 Cosmology for $\zeta=\zeta_1\theta$

To obtain the coefficient of bulk viscosity $\zeta$ is proportional to scalar expansion [66, 67] i.e.

$$\zeta=\zeta_1\theta, \tag{41}$$

where $\zeta_1(\geq 0)$ is a constant. For this case of assumption, equation (28) implies that

$$\dot{\rho}+3(1+\omega)\rho H=27\zeta_1 H^3. \tag{42}$$

Solving equations (31), (34) and (42) the expressions for matter density $\rho$ and the coefficient of bulk viscosity $\zeta$ are take the form as

$$\rho=\frac{9\zeta_1\beta^2}{(1+\omega)}+\frac{54\zeta_1\beta^2}{(3+3\omega-\gamma)(e^{\beta\eta}-1)}+\frac{27\zeta_1\beta^2}{(3+3\omega-2\gamma)(e^{\beta\eta}-1)^2}+\frac{k_3}{\left(e^{\beta\eta}-1\right)^{\frac{3(1+\omega)}{\gamma}}}, \tag{43}$$

$$\zeta=\frac{3\beta\zeta_1}{(1-e^{-\beta\eta})}, \tag{44}$$

Also, the gravitational constant $G$ and cosmological term-$\Lambda$ are as

$$G = \frac{1}{4\pi}\left[\frac{\beta^2\gamma e^{\beta t}}{(e^{\beta t}-1)^2} - \frac{k_1^2}{(e^{\beta t}-1)^{\frac{6}{\gamma}}} - \frac{\alpha^2}{(e^{\beta t}-1)^{\frac{2}{\gamma}}}\right] \times \left[9\zeta_1\beta^2 + \frac{54\zeta_1\beta^2(1+\omega)}{(3+3\omega-\gamma)(e^{\beta t}-1)}\right.$$

$$\left. + \frac{27\zeta_1\beta^2(1+\omega)}{(3+3\omega-2\gamma)(e^{\beta t}-1)^2} + \frac{k_2(1+\omega)}{(e^{\beta t}-1)^{\frac{3(1+\omega)}{\gamma}}} - \frac{9\zeta_1\beta^3 e^{3\beta t}}{(e^{\beta t}-1)^3}\right]^{-1}, \quad (45)$$

$$\Lambda = \left[\frac{2k_1^2}{(e^{\beta t}-1)^{\frac{6}{\gamma}}} + \frac{2\alpha^2}{(e^{\beta t}-1)^{\frac{2}{\gamma}}} - \frac{2\beta^2\gamma e^{\beta t}}{(e^{\beta t}-1)^2}\right] \times \left[\frac{9\zeta_1\beta^2}{(1+\omega)} + \frac{54\zeta_1\beta^2}{(3+3\omega-\gamma)(e^{\beta t}-1)}\right.$$

$$\left. + \frac{27\zeta_1\beta^2}{(3+3\omega-2\gamma)(e^{\beta t}-1)^2} + \frac{k_3}{(e^{\beta t}-1)^{\frac{3(1+\omega)}{\gamma}}}\right] \times \left[9\zeta_1\beta^2 + \frac{54\zeta_1\beta^2(1+\omega)}{(3+3\omega-\gamma)(e^{\beta t}-1)}\right.$$

$$\left. + \frac{27\zeta_1\beta^2(1+\omega)}{(3+3\omega-2\gamma)(e^{\beta t}-1)^2} + \frac{k_2(1+\omega)}{(e^{\beta t}-1)^{\frac{3(1+\omega)}{\gamma}}} - \frac{9\zeta_1\beta^3 e^{3\beta t}}{(e^{\beta t}-1)^3}\right]^{-1}$$

$$+ \frac{3\beta^2 e^{2\beta t}}{(e^{\beta t}-1)^2} - \frac{k_1^2}{(e^{\beta t}-1)^{\frac{6}{\gamma}}} - \frac{3\alpha^2}{(e^{\beta t}-1)^{\frac{2}{\gamma}}}, \quad (46)$$

At $t=0$, the values of $\zeta$, $\rho$, $G$, $\Lambda$ all are diverse. When $t \to \infty$, $p$ become to zero and $\zeta = \zeta_1$. We notice from equation (43) that the matter density decreases with cosmic time (see Fig. 4). From equation (45) we have plotted $G$ versus cosmic time $t$ and it shows negative in nature (see Fig. 5). It is possible that the model with negative gravitational constant represent in early epoch [65]. Also, from equation (46), we have drawn the cosmological term $\Lambda$ against cosmic time $t$ and it seen that $\Lambda$ being large early times and it reduces very small at late time (see Fig. 6)

Expansion scalar $\theta$, shear $\sigma$ and deceleration parameter $q$ for the model are

$$\theta = \frac{3\beta}{(1-e^{-\beta t})}, \quad (47)$$

$$\sigma = \frac{k_1}{\sqrt{3}(1-e^{-\beta n})^{\frac{3}{\gamma}}}, \tag{48}$$

$$q = \frac{\gamma}{e^{\beta n}} - 1, \tag{49}$$

when $t = 0$, the deceleration parameter $q = \gamma - 1$ $(>0)$ and our solution tend to a de Sitter universe. Also, for big values of $t$, we get $q = -1$, which implies the de Sitter universe [68]. Thus our model begins with a decelerating one and expansion transits from decelerating to accelerating phase.

The time $t_q$ for the present model is obtain by putting $q = 0$ in (49)

$$t_q = \frac{\ln \gamma}{\beta \gamma}. \tag{50}$$

For our model,

$$\frac{\sigma}{\theta} = \frac{k_3(1-e^{-\beta n})}{3\sqrt{3}\beta(e^{\beta n}-1)^{\frac{3}{\gamma}}}. \tag{51}$$

For $t \to \infty$, $\frac{\sigma}{\theta} \to 0$, thus our model reduces to isotropy for large values of $t$ and the present age of the universe obtained as

$$\beta n_0 = In\left(\frac{H_0}{H_0 - \beta}\right), \tag{52}$$

where $H_0$ represents the recent value of $H$.

## 6. Concluding remarks

Here, in presence of Bulk viscous fluid source together with time dependent gravitational $G$ and cosmological $\Lambda$ constants are assumed in Bianchi type-$V$ model. The solutions of Einstein's field equations are obtained by using a fractional form of average Hubble parameter suggested by Ellis and Madsen [55]. Also, the coefficient of bulk viscosity is considered to be a power function of matter density i.e. $\zeta = \zeta_0 \rho^l$ and scale of expansion i.e. $\zeta = \zeta_1 \theta$, where $\zeta_0$, $\zeta_1$ and $l$ are constants. The inclusion of bulk viscosity is to produce a change in the cosmic fluid and so exhibits essential change on character of the solution. The describe model becomes to a perfect fluid model for

$\zeta_0 = \zeta_1 = 0$ and it has singularity at $t = 0$. The average scale factor increase with cosmic time and reduces to infinite for big value of $t$. From Figs. 3 and 6, we have seen that the cosmological term $\Lambda$ being very large at earlier times and tend to a genuine cosmological constant at late times, which is supported by the recent observations (Perlmutter et al. [9, 10]; Riess et al. [11, 12]; Garnavich et al. [71, 72]; Schmidt et al. [73]). Additionally, the recent high red-shift type 1a supernovae observations indicated that the universe may be an accelerating one with induced cosmological density through the cosmological $\Lambda$-term. Also, the fractional form of Hubble's parameter depicts an alternative concept to get the explicit solutions of field equation. Hence, it represents a unique picture of the evolution of the universe which starts with a deceleration expansion and expanding with an acceleration at current time (Perlmutter et al. [74]; Knop et al. [75]; Tegmark et al. [76]; Spergel et al. [77]).

Funding Statement: There is no funding agency associated with this manuscript.

Conflict of interest: The author declare there is no conflict of interest.

Data Availability Statement: This manuscript has no associated data.

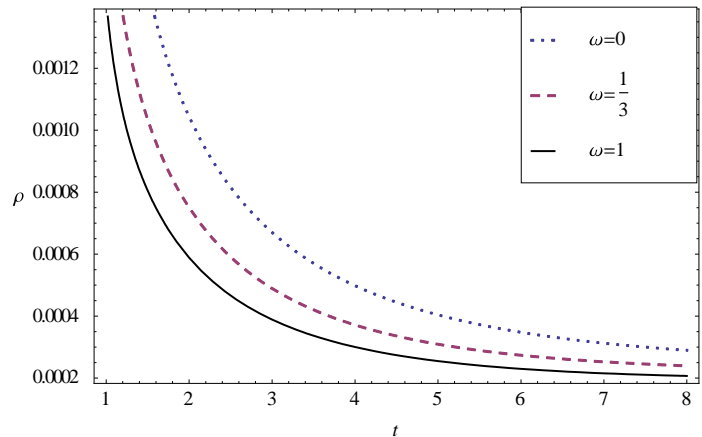

Fig.1 The behavior of matter density $\rho$ versus cosmic time $t$ by considering $\beta = 0.3$, $\gamma = 1.5$, $\zeta_0 = 0.3$, $l = 1.5$ and $k_2 = 0.5$.

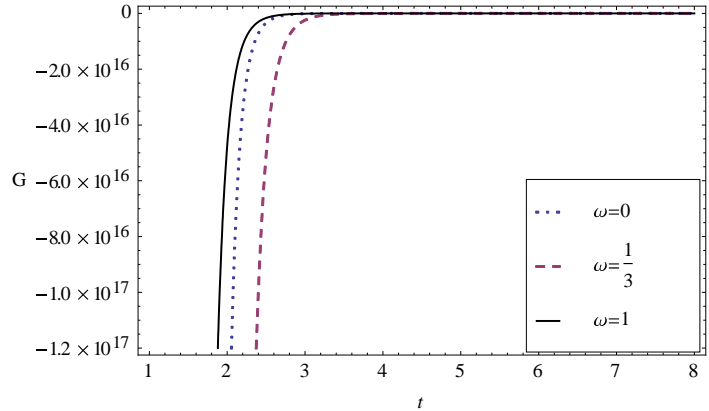

Fig.2 The behavior of gravitational constant $G$ versus cosmic time $t$ by considering $\alpha = 0.4$, $\beta = 0.3$, $\gamma = 1.5$, $\zeta_0 = 0.3$, $l = 1.5$, $k_1 = 0.2$ .and $k_2 = 0.5$.

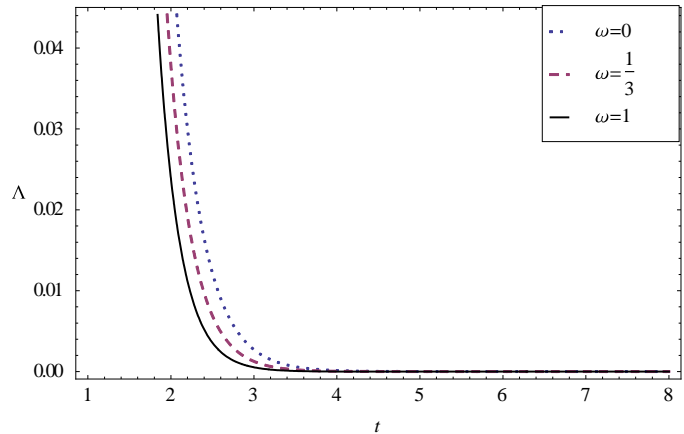

Fig.3 The behavior of Cosmological term-$\Lambda$ versus cosmic time $t$ by considering $\alpha = 0.4$, $\beta = 0.3$, $\gamma = 1.5$, $\zeta_0 = 0.3$, $l = 1.5$, $k_1 = 0.2$ .and $k_2 = 0.5$.

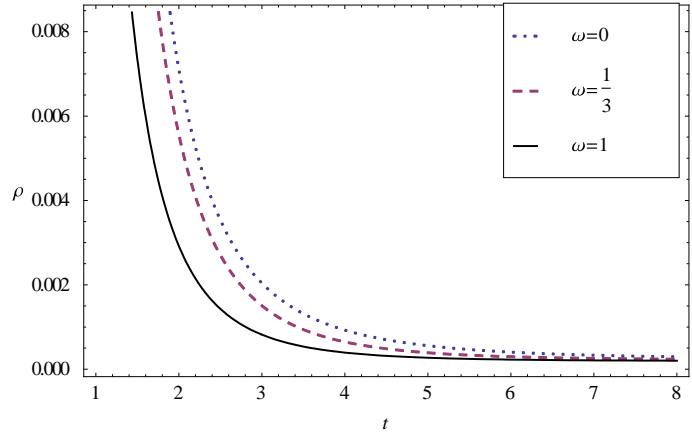

Fig.4 The behavior of matter density $\rho$ versus cosmic time $t$ by considering $\beta = 0.3$, $\gamma = 1.5$, $\varsigma_1 = 0.4$ and $k_3 = 0.4$.

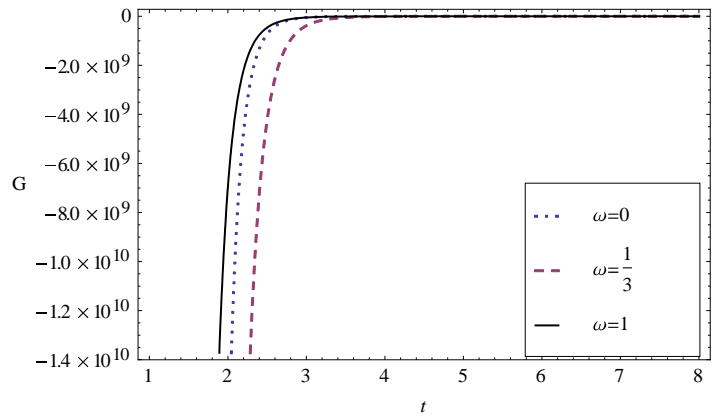

Fig.5 The behavior of gravitational constant $G$ versus cosmic time $t$ by considering $\alpha = 0.4$, $\beta = 0.3$, $\gamma = 1.5$, $\varsigma_1 = 0.4$, $k_1 = 0.2$, and $k_2 = 0.5$.

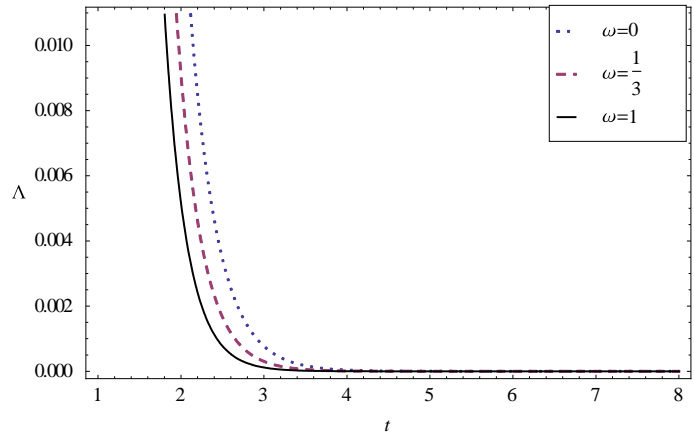

Fig.6 The behavior of Cosmological term-$\Lambda$ versus cosmic time $t$ by considering $\alpha = 0.5$, $\beta = 0.3$, $\gamma = 1.5$, $\zeta_1 = 0.4$ $k_1 = 0.2$, $k_2 = 0.5$ .and $k_3 = 0.4$ .